# Adsorption of a fabric conditioner on cellulose nanocrystals: Synergistic effects of surfactant vesicles and polysaccharides on softness properties

Evdokia K. Oikonomou[1]†, Grazia M.L. Messina[2]†, Laurent Heux[3], Giovanni Marletta[2] and Jean-François Berret[1]*

[1]*Université de Paris, CNRS, Matière et systèmes complexes, 75013 Paris, France*
[2]*Laboratory for Molecular Surface and Nanotechnology (LAMSUN), Department of Chemical Sciences, University of Catania and CSGI, Viale A. Doria 6, 95125, Catania, Italy*
[3]*Centre de recherches sur les macromolécules végétales, BP 53, 38041 Grenoble cedex 9, France*

**Abstract:** Concentrated fabric conditioners are water-based formulations containing 10 wt. % of cationic surfactants that are deposited on textile fibers during the rinse cycle in a washing machine to make them smoother and softer to touch. In topical formulations, the concentration of cationic surfactants is reduced by half, this reduction being compensated by the addition of environment-friendly polysaccharides. Using atomic force microscopy (AFM), quartz crystal microbalance with dissipation (QCM-D) and ellipsometry, the deposition of formulations with or without polysaccharides on model cellulose substrates is studied. We found that 180 nm long cellulose nanocrystals (CNCs) deposited by spin-coating on amorphous silicon dioxide or on quartz crystal sensors provide a good model of cellulose surfaces. QCM-D results reveal strong electrostatic interactions of the surfactant vesicles and polysaccharides with the CNC layer. In the presence of polysaccharides, the adsorbed quantities are less (by 60%) than the sum of the respective amounts of each component, the structure of the interface being however more homogenous and rigid. This outcome suggests that surface techniques coupled with CNC coated substrates are promising for studying interactions of current formulations with cellulose surfaces.

# I – Introduction

In soft matter, soft interfaces are defined as transition zones between two phases, in general a liquid and a solid (Sigel 2017). Such interfaces result from the adsorption of (macro) molecules from the dispersed phase, leading to the formation of a layer a few to several hundred nanometers thick. Prominent examples of soft interfaces are self-assembled monolayers, assemblies of surfactants comprising Langmuir-Blodgett films and micelles, lipid membranes, vesicles and polymer-like chains in various conformations, or combinations thereof. The primary goal of soft interfaces is to modify the surface properties of a solid, and provide materials with controlled, switchable or stimuli-responsive interactions with respect to its (liquid) environment.

One area where soft interfaces are important is that of the deposition of actives on textiles. On cellulose, relevant applications are dealing with laundry and conditioning (Mishra and Tyagi 2007; Murphy 2015). The actives are thereby low molecular-weight compounds such as surfactants and hydrophobic fragrance molecules or polymers. Fabric conditioners (or softeners) are water-based formulations used to impart softness and confer soft handle and smoothness to the clothes. Concentrated softeners contain around 10 wt. % of double tailed cationic surfactants, the surfactants assembling into micron sized vesicles. The vesicles ensure the stability of the dispersion and act as carriers for oils and fragrances onto the cotton. Designed for more than 50 years (Murphy 2015), these formulations are used in the last rinse of the washing cycle during which they adsorb onto cellulose (cotton) fibers. Recently, efforts have been devoted to the design of conditioners based on more environment-friendly materials, with the aim of significantly reducing the quantity of surfactants used, or replacing them by natural compounds. Pertaining to mechanistic softness models, it is accepted that the deposition is driven by electrostatics and the softness effect is due to the formation of lubricating surfactant layers on the fibers, facilitating thereby their mutual contacts and sliding. In most models, the performance of a conditioner is assumed to be correlated with the amount of surfactants adsorbed on the fabrics during the deposition cycle. In the present work, we aim to address this issue and evaluate the deposition of a conditioner formulation on cellulose substrates using surface analytical techniques, such as atomic force microscopy (AFM) (Reimhult et al. 2004), ellipsometry (Stroumpoulis et al. 2006) and quartz crystal microbalance with dissipation monitoring (QCM-D) (Cho et al. 2010; Duarte et al. 2015; Richter et al. 2003; Vermette 2009). QCM-D, in particular, is appropriate as it allows to monitor the kinetics of adsorbed and released masses, whereas the energy dissipation properties provide knowledge on the viscoelasticity of the adsorbed layer (Reviakine et al. 2011). To our knowledge this combination of techniques was not attempted for the evaluation of the deposition of surfactant vesicles on cellulose.

An important factor to consider in the deposition is the type of cellulose substrate that should be used. Regarding the QCM-D technique, crystal quartz resonators coated with microfibrillated cellulose (Henriksson et al. 2007), also referred to cellulose nanofibrils have been designed and are available commercially. To increase the sensitivity in the measurements, it is desirable to increase the specific surface area and/or the electrostatic net charges of cellulose, which can be achieved in using cellulose nanocrystals (CNCs). CNCs are the crystalline part of cellulose fibers and are available as bulk dispersions (Elazzouzi-Hafraoui et al. 2008). Their shapes are mostly rod-like and their sizes around 100 nm. In recent years, CNC coated substrates have received increasing attention with regard to surface techniques, mainly for testing adsorption properties of polymers and proteins (Reid et al. 2017; Villares et al. 2015; Vuoriluoto et al. 2015). Engström *et al.* for instance studied the deposition and adhesion of latex nanoparticles on surfaces of cellulose nanofibrils by QCM-D, and have found that the particle softness, which can be modulated *via* the glass transition temperature of the polymers, was a key parameter for producing strong and tough materials of high nanocellulose content (Engström et al. 2019; Engström et al. 2017). Similar QCM studies have been carried out on CNC substrates for food packaging (Saini et al. 2016), biomimicking wood (Pillai et al. 2014) and enzymatic hydrolysis applications (Lin et al. 2019). Finally, studies combining AFM and QCM have been performed using proteins, antibodies or enzymes (Raghuwanshi and Garnier 2019), showing again the importance of nanocellulose substrates in producing model substrates for material science or biochemistry studies. Despite the high potential of CNC soft interfaces, we are not aware of recent reports dealing with interaction with surfactant vesicles. CNCs and vesicles have been studied in bulk dispersions however. Navon *et al.* (Navon et al. 2017) analyzed the mixtures of CNC-1,2-dioleoyl-*sn*-glycero-3-phosphocholine (DOPC) vesicles in aqueous bulk solutions to understand the interaction between cellulose and a cell membrane mimicking system (Navon et al. 2020). In addition, this group demonstrated by cryo-TEM that the vesicles remain stable during CNC adsorption. On our side, we have shown that the micron sized cationic surfactant vesicles present in fabric conditioners interact strongly with anionic CNCs, forming mixed aggregates and phase separations (Mousseau et al. 2019; Oikonomou et al. 2018; Oikonomou et al. 2017).

Here, we used CNC as a cotton fiber model, assessing the adsorption of topical formulations of double-tailed surfactants and polysaccharides (Oikonomou et al. 2017) . The polysaccharides are cationic guar gum (C-Guar) and a hydroxypropyl guar (HP-Guar), both extracted from the seeds of cyamopsis tetragonalobus plant, and used as additives. According to independent panelists in double-blinded tests as part of an important survey by Solvay, surfactant formulations modified with C-Guar and HP-Guar were shown to display remarkable softness performances as compared to the additive-free benchmark (**Supplementary Information S1**). At the same time the surfactant content of concentrated fabric conditioners could be reduced by about half, from 10.5 wt. % to 6 wt. % (Zhang et al. 2015; Zhang et al. 2016). In this work, we use AFM, QCM-D and ellipsometry to monitor the adsorption of surfactant vesicles and polysaccharides onto model cellulose substrates and to search for a correlation between the structure of the deposited layer and the softness properties.

## II – Materials and Methods
### II-1 - Materials and sample preparation
**Materials**
The surfactant used in the softener formulations (abbreviated TEQ in the following) is the esterquat ethanaminium, 2-hydroxyN,N-bis(2-hydroxyethyl)-N-methyl-esters with saturated and unsaturated C16-18 aliphatic chains. TEQ was provided to us by Solvay and used as received. The counterions associated to the amines are methyl sulfate anions. Small-angle X-Ray scattering performed on TEQ aqueous dispersions ($c$ = 4 wt. %) have revealed the TEQs self-assemble into bilayers of thickness 4.6 nm, whereas wide-angle X-ray scattering disclosed the existence of an hexagonal order in the bilayer at room temperature. This order was found to melt progressively with increasing temperature, indicating a gel-to-fluid transition at the temperature of 60 °C. The ordered surfactant phase thus ensures great stability of the vesicles under the conditions of use. CNCs were prepared according to earlier reports using catalytic and selective oxidation. Briefly, cotton linters provided by Buckeye Cellulose Corporation were hydrolyzed according to the method described by Revol *et al.* (Revol et al. 1992) treating the cellulosic substrate with 65% (w/v) sulfuric acid at 63 °C during 30 min. The suspensions were washed by repeated centrifugations, dialyzed against distilled water until neutrality and sonicated for 4 min with a Branson B-12 sonifier equipped with a 3 mm microtip. The suspensions were then filtered through 8 µm and then 1 µm cellulose nitrate membranes (Whatman). At the end of the process, a 2 wt. % aqueous stock suspension was obtained. The polysaccharides put under scrutiny are a cationic (C-Guar) and a hydroxypropyl (HP-Guar) guar, both synthesized by Solvay. C-Guar and HP-Guar were extracted from the seeds of cyamopsis tetragonalobus plant and their molecular weights were estimated at $0.5 \times 10^6$ g mol$^{-1}$ and $2 \times 10^6$ g mol$^{-1}$ respectively. C-Guar was obtained by introducing positively charged trimethylamino(2-hydroxyl)propyl into the backbone. Water was deionized with a Millipore Milli-Q Water system (resistivity 18.2 MΩ cm). All the products were used without further purification.

**Formulation preparation**

Concentrated softener formulations were prepared using MilliQ filtered water at 60 °C. TEQ surfactants were first melted at 60 °C and added dropwise to water. The final pH was adjusted at 4.5. The dispersion containing guar polysaccharides were produced by first dissolving the guars in water at 60 °C and then adding the melted TEQs. The pH was again adjusted at 4.5 after each step. Mixed dispersions of surfactant vesicles and guars were prepared following the method of continuous variation (Courtois and Berret 2010; Fresnais et al. 2009; Renny et al. 2013) applied to study phase behaviors of oppositely charged colloids (Mousseau and Berret 2018; Mousseau et al. 2018; Mousseau et al. 2016). The following formulations were prepared: TEQ 6 wt. %, TEQ 6 wt. % + C-Guar 0.3 wt. % + HP-Guar 0.3 wt. %, C-Guar 0.3 wt. % + HP-Guar 0.3 wt. %. These dispersions were then diluted 60 times, leading to the three dispersions: TEQ 0.1 wt. %, TEQ 0.1 wt. % + Guars 0.01 wt. %, Guars 0.01 wt. % (HP-Guar 0.005 wt. %, C-Guar 0.005 wt. %).

**Cellulose coated substrates**

For the deposition of surfactants and polysaccharides on cellulose, two different substrates were investigated. We first considered the Q-Sense sensor (QSX 334, Biolin Scientific) which is described as a model surface of native cellulosic fibers. The type of cellulose used is microfibrilated cellulose, also referred to as nanofibrillar cellulose in literature (Henriksson et al. 2007; Pääkkö et al. 2007). Its adhesion to the quartz crystal is mediated *via* a poly(ethylene imine) layer. As recommended by the supplier, Q-Sense sensors were gently rinsed before use with water and dried with nitrogen gas. An AFM image of the microfibrilated cellulose is provided in **Supplementary Information S2**. The second type of substrate was achieved through spincoating technique of a cellulose nanocrystal dispersion on model amorphous silicon dioxide or quartz surfaces. Spincoating was selected as particularly suitable for obtaining reasonably uniform coatings. Several protocols have been tested to evaluate the optimal coating conditions, by modifying the quantity of CNC dispersions, the spinning speed and the spinning time. Good results have been obtained by depositing and drying, onto amorphous silicon dioxide or quartz substrates, a single dose of 100 µl of a 1.4 wt. % CNC dispersion in water with a spinning speed of 3500 rpm for 30

s. Once spincoated, the solvent was allowed to evaporate and the coated substrates were stored under atmospheric conditions.

## II-2 Methods

**Dynamic Light scattering (DLS) and zeta potential**

The scattering intensity $I_S$ and hydrodynamic diameter $D_H$ were measured using the Zetasizer Nano ZS spectrometer (Malvern Instruments, Worcestershore, UK). A 4 mW He−Ne laser beam ($\lambda$ = 633 nm) is used to illuminate the sample dispersion, and the scattered intensity is collected at a scattering angle of 173°. The second-order autocorrelation function $g^{(2)}(t)$ is analyzed using the cumulant and CONTIN algorithms to determine the average diffusion coefficient $D_C$ of the scatterers. The hydrodynamic diameter is then calculated according to the Stokes-Einstein relation, $D_H = k_B T / 3\pi\eta D_C$, where $k_B$ is the Boltzmann constant, $T$ the temperature and $\eta$ the solvent viscosity. Laser Doppler velocimetry using the phase analysis light scattering mode and detection at an angle of 16° was used for electrokinetic measurements of electrophoretic mobility and zeta potential with the Zetasizer Nano ZS equipment (Malvern Instruments, UK). DLS and Zeta potential was measured in triplicate after a 120 s equilibration at 25 °C.

**Cryo-transmission electron microscopy (cryo-TEM)**

For cryo-TEM, few microliters of dispersion (concentration 1 wt. % for TEQ and 0.1 wt. % for CNCs) were deposited on a lacey carbon coated 200 mesh (Ted Pella Inc.). The drop was blotted with a filter paper using a FEI VitrobotTM freeze plunger. The grid was then quenched rapidly in liquid ethane to avoid crystallization and later cooled with liquid nitrogen. It was then transferred into the vacuum column of a JEOL 1400 TEM microscope (120 kV) where it was maintained at liquid nitrogen temperature thanks to a cryo-holder (Gatan). The magnification was comprised between 3000× and 40000×, and images were recorded with an 2k×2k Ultrascan camera (Gatan). Images were digitized and treated by the ImageJ software and plugins (http://rsbweb.nih.gov/ij/).

**Quartz crystal microbalance with dissipation monitoring**

Measurements of adsorption kinetics were performed by using a QCM-D instrument (Q-Sense AB, Sweden) with AT-cut gold crystals sensors. The measurement chamber was operating in water at 25 ± 0.1° C. The simultaneous measurements of frequency $f$, and energy dissipation, $D$, were performed for the fundamental resonance frequency ($n$ = 1, *i.e.* $f$ = 5 MHz) and the six overtones (n = 3, 5, 7, 9, 11 and 13 corresponding to $f$ = 15, 25, 35, 45, 55 and 65 MHz, respectively). The resolution in $f$ and $D$ is ± 0.1 Hz and 1×10⁻⁷, respectively. Regarding the analysis and discussion of QCM-D data, we will focus on the third overtone $\Delta f_3/3$ because, in contrast to higher harmonics it probes the bulk solution above the sensor (Feiler et al. 2007; Giamblanco et al. 2017). In the cases of rigid, evenly distributed and sufficiently thin adsorbed layers, the frequency to mass conversion is assumed to be given by Sauerbrey equation (Sauerbrey 1959) : $\Delta m = -C\, \Delta f_n/n$, where $f$ is the decrease in resonant frequency, $\Delta m$ is the mass uptake at the sensor surface, $C$ is a constant depending on the intrinsic properties of quartz slab (in our case $C$ = 17.7 ng cm⁻² Hz⁻¹ at $f$ = 5 MHz) and $n$ (= 1, 3, …) is the overtone number (Rodahl et al. 1997). Amorphous silicon dioxide substrates and quartz crystal sensors prepared for the surface experiments were obtained by spincoating a 1.4 wt.% CNC aqueous dispersion, followed by the evaporation of the solvent. Each QCM-D experiment started with the sensor running in water (outgassed with 30 min of sonication), then the addition of vesicles and softener and, after 30 minutes the exchange of the solution being measured with water, to check both desorption and stability of the adsorbed layer. All the experiments were performed in water at 25 °C and the flow rate was 150 µL min$^{-1}$. Three replicas for each experiment were performed for data reproducibility.

**Atomic force microscopy**

Topographical images were obtained using a commercial Nanoscope IIIa Multimode AFM Instrument (Digital Instruments, Santa Barbara, CA, USA). The device was equipped with a <J> calibrated scanner using grating manufacturers. All samples were observed in tapping mode using 0.5 – 2 Ω cm phosphorous n-doped silicon tips mounted on cantilevers with a nominal force constant of 40 N m$^{-1}$ and a resonant frequency of 300 kHz. The applied force was varied over a range from several nanoNewtons up to tens of nanoNewtons in contact mode. Measurements in liquid have been performed using liquid cell and Milli-Q ultrapure water has be injected with a syringe. The surface was analyzed using triangular silicon nitride tip having a nominal force constant of 0.12 N m$^{-1}$ and a resonant frequency of 20 KHz. The glass surfaces were examined *ex situ* after gently dried. Image analysis was carried out using DI software v4.23r6. The images were flattened to remove background slopes. Film roughness was measured on tapping mode images obtained at a scan speed of 1 Hz over scanned areas of 500×500 nm$^2$ for a minimum of three separate images for each sample obtained from different regions.

**Spectroscopic Ellipsometry**

The coated samples were characterized at room temperature in M2000 ellipsometer (rotating compensator ellipsometer, J.A. Woollam Co. Inc.) in a wavelength range of 400 – 1600 nm with a step size approximately 1.5 nm at two different incidence angles of 65° and 75°. The acquired data were analyzed using WVASE32 software module. This technique measures the ratio of the Fresnel reflection coefficients $R_p$ and $R_s$ obtained respectively for p-polarized and s-polarized light reflected from the surface sample. Measured values, in the form of the ellipsometric angles $\Psi$ and $\Delta$, are described from the following equation: $P = R_p/R_s = tan\Psi e^{i\Delta}$. VASE is an indirect method used to extract the optical constants as well as the film layer thickness and it is based on the fitting of experimental data with the modelled data and generally the quality of fit is given by an estimated value which is a positive quantity and tends to zero when the experimental data approaches or exactly matches the calculated data. Cauchy model was used to match the acquired data with the generated theoretical data

# III - Results and discussion

### III.1 – Surfactant, polysaccharides and cellulose nanocrystals

Fig. 1a displays the molecular structure of the TEQ surfactant and a representative image of a 1 wt. % dispersion studied by cryo-TEM. The image shows uni- and multivesicular vesicles of size between 50 nm and 500 nm in a crowded environment. Morphologies characteristic for double-tailed surfactant self-asssembly, including deflated vesicles and doublets are also observed. At the concentration studied (0.1 wt. %) and at the concentration of use (0.024 wt. %), the TEQ vesicles remain stable, the proportion of multivesicular structures compared to the unilamellar ones reversing under dilution. Electrophoretic mobility measurements on dilute dispersions resulted in zeta potential $\zeta$ = +65 mV, indicating strongly cationic vesicles. The cellulose nanocrystals were investigated *via* electron microscopy and light scattering. Fig. 1b show a cryo-TEM image of CNCs in the form of thin anisotropic fibers (Oikonomou et al. 2018). Their average length, width and thickness were derived from image analysis and were found equal to 180 ± 30 nm, 17 ± 4 nm and 7 ± 2 nm, respectively. Using light scattering, CNC dispersions (pH 4.5) prepared at 0.1 wt. % display a single relaxation mode in the autocorrelation function $g^{(2)}(t)$. This function is seen to decrease rapidly above delay times of $10^3$ μs, suggesting the absence of CNC aggregates. The hydrodynamic size retrieved by DLS reveals an intensity distribution centered around 120 nm (Fig. 1b), in agreement with the size obtained from electron microscopy. Electrophoretic mobility experiments resulted in zeta potential $\zeta$ = -38 mV and negatively charged CNCs. Figs. 1c and 1d show the molecular structures of C-Guar and HP-Guar respectively, together with and the $g^{(2)}(t)$-functions obtained by DLS from dispersions at 0.02 wt. % (after filtration at 0.2 μm). The intensity distributions in the insets reveal a single peak around 200 nm for C-Guar and a double peak for HP-Guar at 50 and 350 nm (Oikonomou et al. 2018). These hydrodynamic diameters are larger than those expected from polysaccharides of molecular weight $M_W$ = 0.5 and $2 \times 10^6$ g mol$^{-1}$, as provided by Solvay, suggesting that the chains are likely associated in water, forming hydrocolloid particles (Rubinstein and Colby 2010). Electrophoretic mobility measurements on dilute guar dispersions give zeta potential of +30 mV and 0 mV respectively. The dynamic light scattering and $\zeta$-potential data are summarized in Table I.

**Table I**: *Hydrodynamic diameter ($D_H$) and $\zeta$-potential determined for TEQ surfactant vesicle at 0.1 wt. % (Oikonomou et al. 2017), guar dispersions at 0.02 wt. % and cellulose nanocrystals at 0.1 wt. %. The dispersity indexes (pdi) for the 4 compounds were in the range 0.2 – 0.3 and the uncertainty in $\zeta$-potentials found at ± 5 mV.*

| Compounds | Abbreviation | $D_H$ (nm) | $\zeta$-potential (mV) |
|---|---|---|---|
| Surfactant vesicles* | TEQ | 620 | + 65 |
| Cationic Guar | C-Guar | 200 | + 30 |
| Hydroxypropyl Guar | HP-Guar | 50, 350 | 0 |
| Cellulose nanocrystals | CNC | 120 | -38 |

*(Oikonomou et al. 2017)

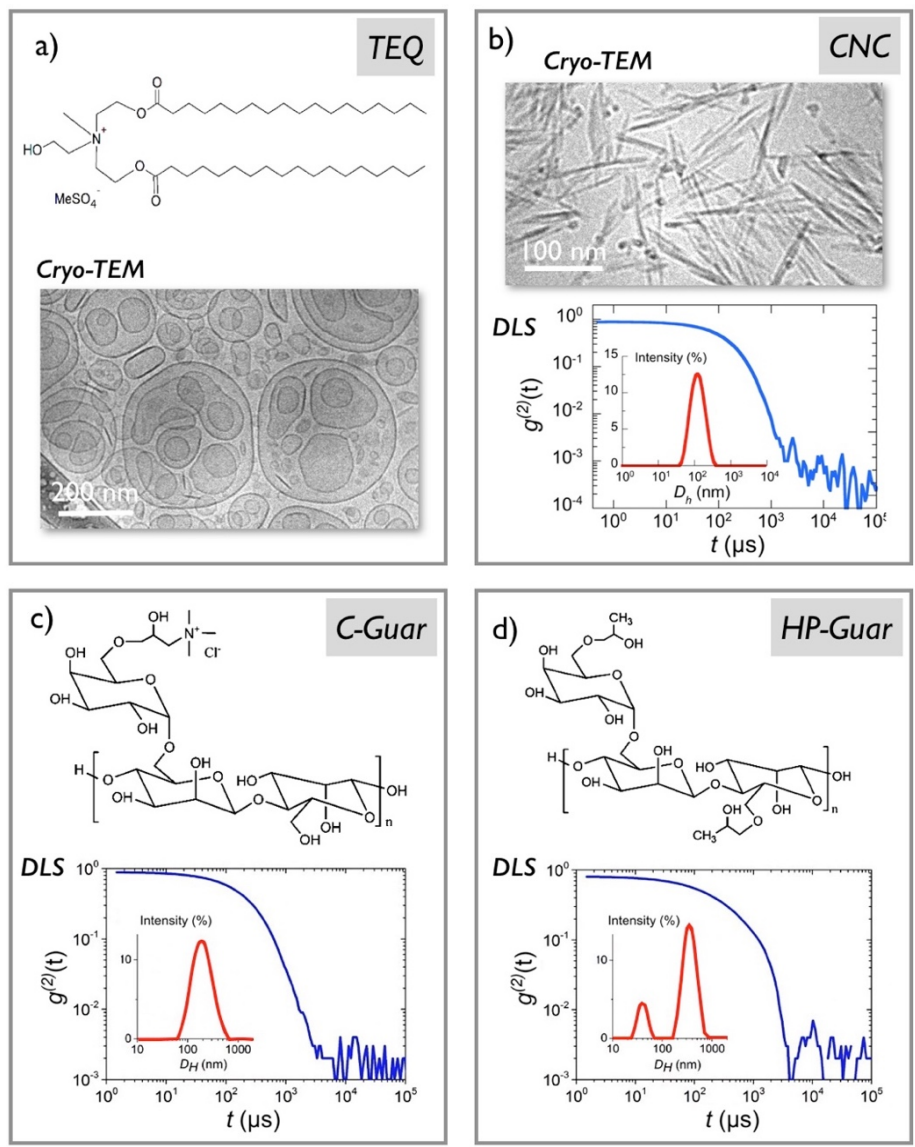

***Figure 1: a)*** *TEQ double-tailed surfactant chemical structure and cryo-transmission electron microscopy image (cryo-TEM) of 1 wt. % aqueous dispersion.* ***b)*** *Cryo-TEM image and dynamic light scattering (DLS) characterization of 0.1 wt. % CNC dispersion (Oikonomou et al. 2018).* ***c)*** *and* ***d)*** *Chemical structure and dynamic light scattering characterization of the cationic and hydroxypropyl modified guars, respectively (Oikonomou et al. 2018).*

### III.2 – Cellulose nanocrystals on silica substrates

Fig. 2a displays AFM tapping mode images in air of cellulose nanocrystals spincoated onto a quartz crystal. It can be seen that a uniform coverage of the substrate is achieved, consisting of densely packed CNCs. The average thickness and roughness of the CNC layer were determined by ellipsometry at 34 nm and 3.9 nm, respectively (**Supplementary Information S3**). These data, compared with those obtained from cryo-TEM experiments in Fig. 1b, suggest that the layers consist primarily of side-on lying crystallites, arranged with the longitudinal axes parallel to the surface (Navon et al. 2020). It should be noted that the coating locally exhibits correlations in the CNC orientations. These correlations, which extend over a length scale of a few microns, attest of strong steric interactions between the nanocrystals in the dry state. To assess the CNC layer stability, an important step in monitoring repeatible QCM-D testing, the previous coated substrate was rinsed with Milli-Q ultrapure water and studied in air with AFM, again after solvent evaporation. Fig. 2b shows the AFM images in air of the CNC after rinsing. It can be seen that the rinsing treatment tends to randomize the CNC orientations, most likely through capillary forces, but it does not modifty the uniformity or the thickness of the adsorbed layer. AFM measurements were then conducted in liquid onto a CNC coating layer immersed in water. The AFM images in Fig. 2c have characteristics similar to those found in the previous conditions, confirming the high stability of the CNC layer under wet and dry conditions, and its relevance for QCM-D. Fig. 2d provides a schematical representation of the CNC layer obtained using the spincoating technique described previously.

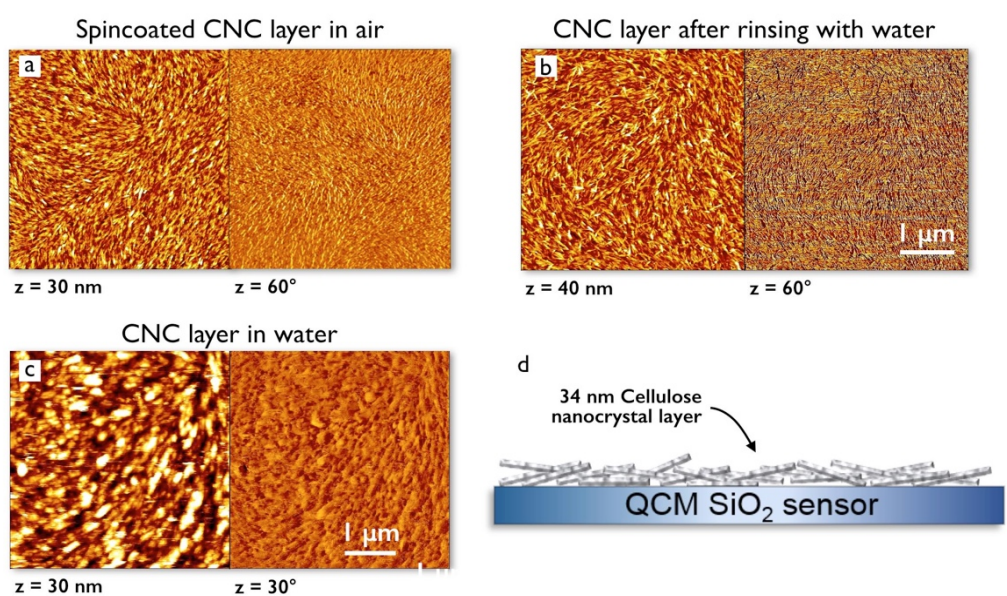

***Figure 2:*** *AFM tapping mode height (left panel) and phase (right panel) images of cellulose nanocrystals deposited on a quartz crystal sensor in different preparation conditions. **a)** In air after spin-coating; **b)** in air after rinsing the substrate shown in **a)** with Milli-Q ultrapure water; **c)** in water. The z-values in the height images represent the vertical distance between the tip and the sample, whereas the z-values in the phase images denote the phase shifts expressed as an angle in degrees and converted in bright and dark regions in the phase images. d) Schematic representation of the CNC layer.*

### III.3 - Adsorption of surfactants and guar polysaccharides on cellulose using QCM-D

To study the interaction of the cationic double tailed surfactants (TEQ) with cotton fibers, QCM-D experiments have been carried out with two types of cellulose substrates: the commercially available Q-Sense sensor coated with microfibrilated cellulose, and a substrate obtained by spin-coating a CNC dispersion on a quartz crystal sensor. For QCM-D, the same procedure was applied with both surfaces.

*Adsorption on microfibrilated cellulose Q-Sense sensor*

QCM-D data obtained with a microfibrilated cellulose coated sensor are presented in **Supplementary Information S2**. Experiments were first conducted in triplicate with TEQ and guar dispersions at 0.1 and 0.01 wt. % respectively, but the QCM-D signal was too low to be exploited quantitatively. In a second attempt, the concentrations were increased tenfold, to 1 and 0.1 wt. %, leading to the adsorption kinetics shown in **Figs. S2c and S2d**. For TEQ alone, the frequency shift exhibits a decrease followed by a saturation at $\Delta f_3/3 = -6$ Hz, while for guars this decrease reaches -15 Hz. These $\Delta f_3/3$ values are relatively low, indicating a weak deposition of each of the two components studied TEQ and Guars, the densities of adsorbed materials being around 200 ng cm$^{-2}$. When the two compounds were combined in the same sample, the adsorption level remains of the same order ($\Delta f_3/3 = -16$ Hz). It should be noted that the concentrations used here are much higher than those used in real conditions (0.024 wt. %), and for this reason these substrates were not considered further as model for the assessment of actives on cotton. A possible reason for such a weak adsorption could stem from the low density of nanofibers and/or electrostatic charges at the surface sensor. In particular, it should be noted that independent measurements of the zeta potential on these dispersions of nanofibrils gave values of -10 mV (Stenstad et al. 2008), against -34 mV for the CNCs.

*Adsorption on cellulose nanocrystal (CNC) coated sensor*
In this section, the QCM-D results obtained using standard quartz sensors preliminarily coated *ex situ* with a 34 nm CNC layer. Fig. 3a displays the adsorption profiles of a 0.1 wt. % TEQ dispersion obtained at room temperature. In the figure, the traces corresponding to the different overtones ($n$ = 3, 5, 7, 9 and 11) and the related dissipations, $\Delta D_n$ are reported. Upon injection (arrow at $t = 10$ min), the frequency exhibits a rapid decrease and then a saturation for all overtones. In Fig. 3a, the frequency shift measured for the third overtone is large, at $\Delta f_3/3 = -225$ Hz. Similarly, $\Delta D_3$ increases rapidly and reaches a plateau at $55 \times 10^{-6}$. This adsorption kinetics, and in particular the adsorption saturation behavior, are consistent with those reported by QCM-D upon adsorption of close-packed layer of intact lipid vesicles (Keller and Kasemo 1998; Richter et al. 2003; Serro et al. 2012; Viitala et al. 2007). Rinsing with water (arrow at $t = 40$ min) has no effect on the frequency shifts, suggesting that the vesicles are irreversibly adsorbed on the surface. These results confirm findings obtained on vesicle/CNC interaction in the bulk phase and the persistence of the vesicular structures upon adsorption on nano- and microcellulose fibers (see **Supplementary Information S4** and Ref. (Oikonomou et al. 2018) for more details). A schematic representation of the adsorption mechanism on the CNC coated crystals is depicted in Fig. 3a.

The role of the guar polysaccharides was then examined by performing two complementary experiments involving *i)* the sequential adsorption of a C-guar and HP-guar mixture (0.005 wt. % each), followed by the adsorption TEQ surfactants at concentration 0.1 wt. % (Fig .3b), and *ii)* the simultaneous adsorption from a formulation containing TEQ surfactants at 0.1 wt. % and guars at 0.01 wt. % (Fig. 3c). Fig. 3b shows that guar molecules are at first strongly adsorbed at the sensing substrate, with an immediate adsorption peak reaching $\Delta f_3/3 = -83$ Hz and a large dissipation ($\Delta D_3 = 18 \times 10^{-6}$), immediately followed by a partial mass loss up to a plateau at $\Delta f_3/3 = -49$ Hz and a dissipation decrease. The subsequent injection of the TEQ surfactant dispersion ($t > 57$ min in Fig. 3b) produces a small and reversible adsorption of nonruptured vesicles, suggesting that the precoating of the CNC layer by the guars prevented further immobilization of the vesicles. Pertaining to the data in Fig. 3b, it is interesting to note that the overall process reminds the characteristic two-step mechanism observed for lipid vesicles interacting with quartz crystal sensors (Reimhult et al. 2003; Reviakine et al. 2012). This process involves first the adsorption of vesicles, immediately followed by their spontaneous rupture and leading to the formation of a supported lipid bilayer and to a significant loss of the water initially trapped in the lipid structures. To justify this analogy, one has to recall that the guar dispersed in aqueous medium behave as hydrocolloids (*i.e.* highly solvated crosslinked polymer networks) rather than ideal linear polymer chains in good solvent. For the cationic and hydroxypropyl guars (Fig. 1), the hydrodynamic diameters derived from DLS are 200 and 350 nm respectively, that is much larger than the expected diameters (estimated at 40 and 90 nm) retrieved from their molecular weights $0.5 \times 10^6$ g mol$^{-1}$ and $2 \times 10^6$ g mol$^{-1}$ respectively. The frequency shift variation in Fig. 3b may then be explained in terms of the initially strong adsorption of hydrocolloids, carrying their solvation water on the CNC surface and producing the observed large frequency shift. This peak is then followed by the release of the water, leaving the polymer chains tightly bound to the fibers. For the cationic guar, this interaction is driven by electrostatics, as shown in bulk studies (Oikonomou et al. 2018). This picture is further supported by results obtained on cationic polyelectrolytes adsorbed onto CNC coated substrates, in which the frequency shift variation has been attributed to the water release from the interface upon charge neutralization (Vuoriluoto et al. 2015).

The narrow distribution of the frequency overtones, along with the lack of further desorption upon rinsing (arrow at $t$ = 40 min), confirms the existence of strong interactions between guars and CNCs, and suggests that following their adsorption, the guars are likely to interpenetrate the cellulose nanocrystal network, as illustrated in Fig. 3b. This hypothesis is also supported by the peculiar decrease of the dissipation to negative values, *i.e.*, below the initial dissipation baseline. To account for this effect, it must be reminded that in Fig. 3 the prior-to-injection baselines in $\Delta f$ and in $\Delta D$ represent the reference states that was previously characterized in pure water. These baselines reflect the mechanical properties of the CNC layer, *i.e.* in particular the connectivity of the CNC network and the related CNC layer organization. The seemingly negative dissipation may be then understood in terms of the binding effect induced by the positively charged guar molecules connecting the negative cellulose nanocrystals in the layer. These interactions produce a global reorganization of the polymer/CNC interface, driving the dissipation to negative values with respect to the reference state. Effects implying change of the substrate dissipation to negative values have been also reported for bacterial membranes disrupted by aminoglycosides (Joshi et al. 2015), phospholipid layers damaged by membrane-active peptides (John et al. 2018), and disruption of layers of fatty acid friction modifiers (Zachariah et al. 2019).

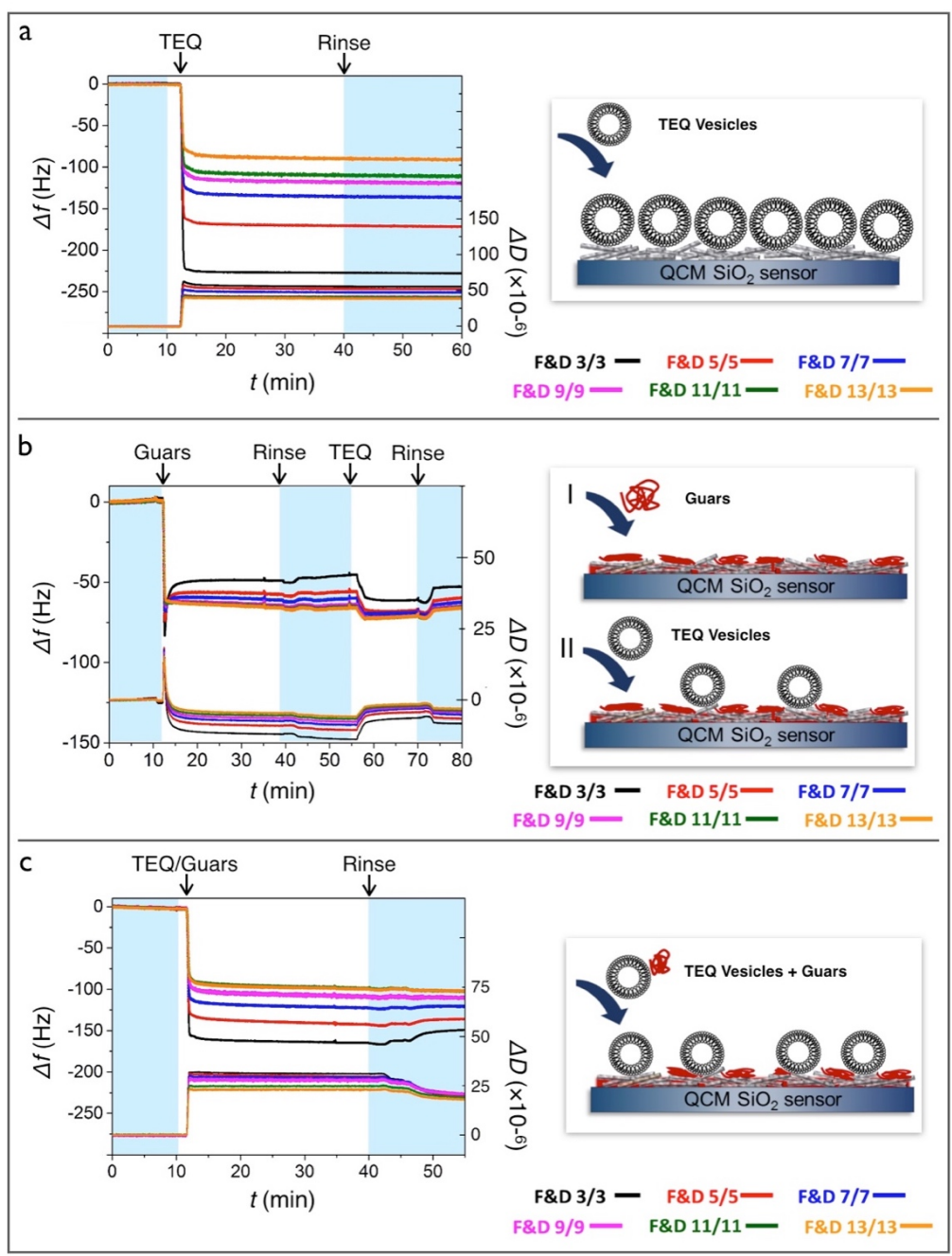

***Figure 3***: *a) Real time binding curves and dissipation obtained by QCM-D during the adsorption of a) TEQ surfactants at 0.1 wt. %, b) a mixture of C-guar and HP-guar at 0.005 wt. % each followed by TEQ surfactants at 0.1 wt. % and c) a formulation containing TEQ surfactants at 0.1 wt. % and C-guar and HP-guar (each at 0.005 wt. %). In these measurements, the quartz sensors were previously coated with cellulose nanocrystals. In each panel, the first arrow at t = 10 min denotes the time at which the sample is injected. The second arrow denotes the time at which Milli-Q ultrapure water is introduced for rinsing. Illustrations of the adsorption processes involved are provided on the sketches on the right.*

To highlight the role of guars on the deposition, we then examined a formulation that contains TEQ vesicles at 0.1 wt. % and guars at 0.01 wt. %. As displayed in Fig. 3c, the injection of the mixed formulation resulted in a large frequency shift at $\Delta f_3/3 = -162$ Hz and in a dissipation of $\Delta D_3 = 31 \times 10^{-6}$. It must be emphasized here that the frequency shift is stable over time, also after the rinsing treatment, showing thus that no mass loss is involved in the vesicle and guar adsorption onto the CNC layer. This in turn suggests that vesicles do not break during the adsorption step, and remain firmly bound to the CNCs. On the other hand, the fact that the overall TEQ surfactant/guars adsorbed mass is about 60% lower than the sum of TEQ and guars taken separately indicates that the adsorbed the guar biomolecules interfere with the vesicle adsorption, screening the adsorption sites on the CNC layer. This picture is also supported by the observation that there is a significant spread of the overtones for TEQ surfactants/guars. This later outcome suggests that a lower amount of mobile vesicles are adsorbed together with random guar aggregates, as pictured in Fig. 3b. The QCM-D experiments in Fig. 3 were repeated and provide similar results (**Supplementary Information S5**).

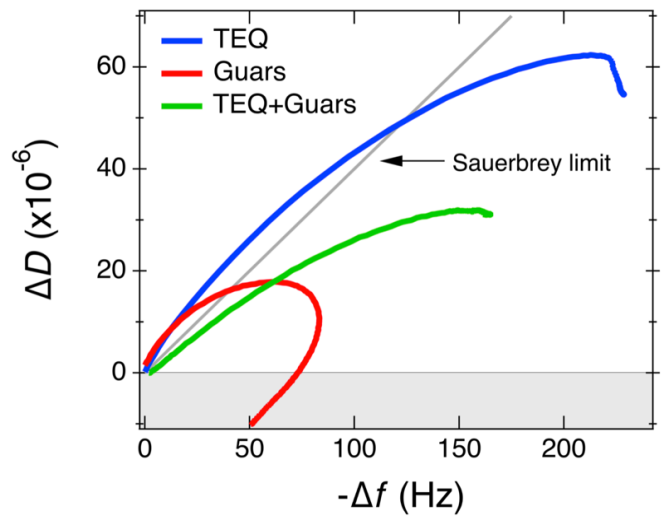

*Figure 4*: *Plot of the dissipation $\Delta D_3$ as a function of the frequency shift, $-\Delta f_3/3$, during the adsorption of TEQ surfactant vesicles, guars and mixture of surfactants and guars. The QCM-D data are those of the first 30 minutes shown in Fig. 3. The straight line represents the Sauerbrey limit (slope $4 \times 10^{-7}$ Hz$^{-1}$) valid for homogeneous and rigid films (Dixon 2008; Reviakine et al. 2011).*

Fig. 4 shows the plot of $\Delta D_3$ *versus* $-\Delta f_3/3$ during the adsorption of TEQ vesicles, guars and mixture of vesicles and guars on CNC coated substrates. The data are those of Fig. 3a, 3b and 3c respectively for which only the first initial 30 minutes have been retained. These $\Delta D$-$\Delta f$ plots allow to overcome the temporal effects pertaining to the deposition kinetics and to directly compare the ratio between $\Delta D$ and $-\Delta f$, that is, the induced energy dissipation per coupled unit mass. A low $\Delta D/(-\Delta f)$ value indicates a mass addition without significant dissipation increase, which is characteristic of a fairly rigid layer, whereas a large $\Delta D/(-\Delta f)$ value signals a soft and dissipative film. The figure also shows the Sauerbrey limit: $\Delta D/(-\Delta f) = 4\times 10^{-7}$ Hz$^{-1}$ (straight line in grey) that is typical of homogeneous and rigid films. Data in Fig. 4 highlight that for TEQ vesicles and guars, the $\Delta D/(-\Delta f)$-responses lie above the Sauerbrey limit at the adsorption onset. The measured responses are characterized by a large dissipation, and both $\Delta f$ and $\Delta D$ display overtones dependencies that are typical for acoustically non-rigid films, probably due to the vesicle intrinsic properties and the nature of their attachment to the surface. In contrast, mixture of vesicles and guars display low $\Delta D/(-\Delta f)$ ratios over the entire time range, indicating an acoustically rigid layer likely due to the presence of polysaccharides adsorbed on CNC and the presence of only few vesicles on surface (Cho et al. 2010). A change in structure of the water-cellulose interfacial layer induced by the guars could explain the enhanced performances found for these novel formulations (**Supplementary Information S1**).

### III.4 – Deposition thickness in the dry state studied by ellipsometry

An important question related mechanics involved in the softness effect concerns the drying step, in which the surfactant vesicle adsorbed on the fabrics transforms into a dense film through water evaporation. Here we study with AFM and ellipsometry the topographical features of a CNC coated adsorbing surfaces, with and without TEQ surfactant or guars. In the dry state, the CNC layer spincoated on a quartz crystal sensor shows a densely packed in CNCs associated with a dry average thickness $h_{SE} = 34.3 \pm 0.1$ nm and with a root mean square roughness $R_q = 3.9 \pm 0.3$ nm (Fig. 5a). The roughness value reflects the CNC layer porosity and it is related also to the dimensions of the cellulose nanocrystallites. The CNC substrate topography, after the adsorption and drying of a 0.1 wt. % dispersion in water of TEQ vesicles is shown in Fig. 5b. The AFM images, in this case, show the appearance of large flake-like structures, randomly distributed onto the CNC coating. The overall process is associated to the net increase in $h_{SE}$ from $34.3 \pm 0.1$ nm to $50.3 \pm 0.1$ nm, roughly corresponding to 3.5 times the thickness of TEQ bilayers determined by small-angle X-Ray scattering (Oikonomou et al. 2017). Accordingly, the roughness is almost doubled, from $R_q = 3.9 \pm 0.3$ nm to $7.6 \pm 0.3$ nm. Fig. 5c, finally, shows the effects of the adsorption and drying of a mixture of TEQ surfactant at 0.1 wt. % and guars at 0.01 wt. % on the CNC layer. The topography is somewhat intermediate between the pristine one and that found with TEQ alone (Table II). Furthermore, the AFM image suggests a large reduction of the CNC orientational correlations, which are attributed to the partial penetration of the guars inside the CNCs. The 30% decrease in layer thickness with guars (from 16.0 nm to 11.1 nm, Table II) is in line with the decrease of deposited materials found with QCM-D in wet conditions.

**Table II** : *Root mean square ($R_q$), average ($R_a$) and maximum ($R_{Max}$) roughness obtained from AFM image analysis. $h_{SE}$ denotes the average layer thickness determined from spectroscopic ellipsometry.*

| Sample | Roughness (nm) | | | Layer Thickness (nm) |
|---|---|---|---|---|
| | $R_q$ | $R_a$ | $R_{Max}$ | $h_{SE}$ |
| CNC 1.4 wt. % | 3.9 ± 0.3 | 3.1 ± 0.3 | 36.1 ± 2.8 | 34.2 ± 0.1 |
| CNC 1.4 wt. % + TEQ wt. 0.1% | 7.6 ± 0.3 | 6.0 ± 0.3 | 51.7 ± 6.6 | 50.3 ± 0.1 |
| CNC wt. 1.4% + TEQ wt. 0.1% + Guars 0.01 wt. % | 7.3 ± 1.2 | 5.8 ± 0.9 | 47.7 ± 3.4 | 45.4 ± 0.1 |

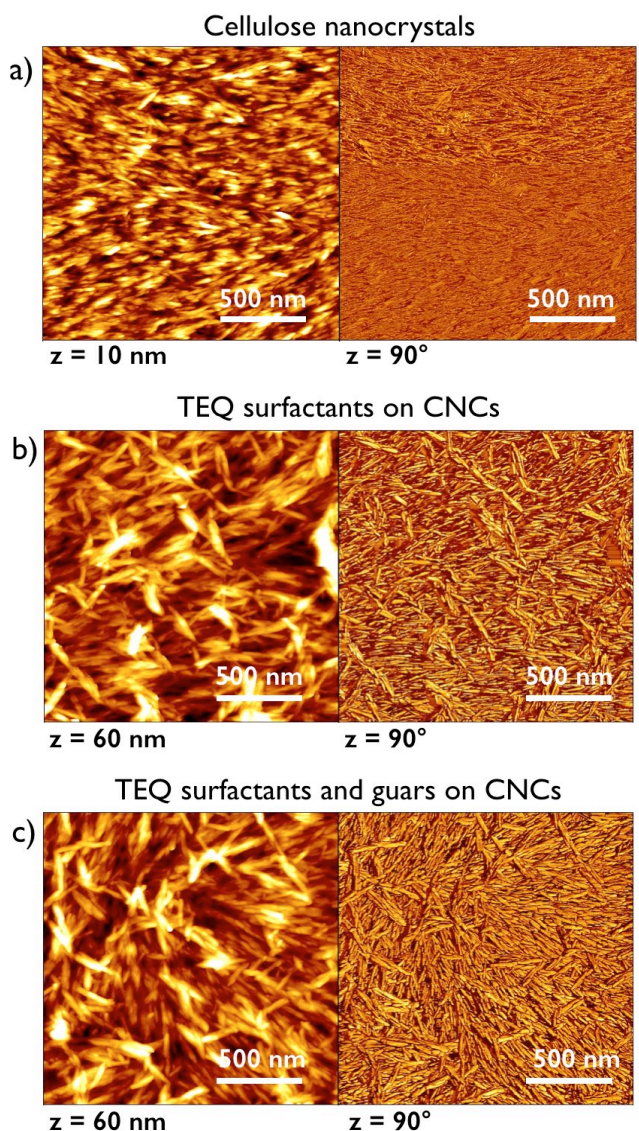

*Figure 5*: AFM tapping mode height (left panel) and phase (right panel) images of **a)** cellulose nanocrystals obtained by spin coating on a quartz crystal sensor; **b)** TEQ surfactants at concentration 0.1 wt. % adsorbed on the substrate shown in **a)**; **c)** a softener formulation containing TEQ surfactants 0.1 wt. % and guars 0.01 wt. % adsorbed on the substrate shown in **a)**. The z-values in the height images represent the vertical distance between the tip and the sample, whereas the z-values in the phase images denote the phase shifts expressed as an angle in degrees and converted in bright and dark regions in the phase images. The images were recorded after adsorption and drying. The scale of the AFM pictures is 2×2 µm$^2$.

# IV - Conclusion

We report the deposition of surfactants and polysaccharides present in topical fabric conditioners on model cellulose substrates. The conditioner performances for these novel formulations were assessed in relation to several properties, including their softness effect and wettability. In the present context, the emphasis is put on the relationship between the softness effect and the deposition of actives on cotton surfaces. To tackle this issue, we use a combination of surface techniques, such as atomic force microscopy, quartz crystal micro-balance and ellipsometry to study the deposition of formulations with or without guars on amorphous silicon dioxide substrates or on quartz crystal sensors. To model cotton fiber surfaces, these substrates were previously coated with cellulose nanofibers, either microfibrilated cellulose or CNCs. The data obtained with CNCs are the most interesting, as they display good reproducibility under wet and dry conditions and great sensitivity to the deposition of cationic vesicles and polymers. The QCM-D technique reveals strong electrostatic interactions between the surfactant vesicles, or the cationic guars and the CNC layer. Upon successive rinsing, we do not observe any loss of mass, indicating a strong adhesion onto the cellulose nanocrystals. An interesting point is that in the wet state the vesicles remain intact during the deposition, while during evaporation of the solvent and subsequent drying the vesicles rupture and produce a uniform film. The fact that vesicles are adsorbed on CNCs without modification of their structure is supported by additional experimental results (Oikonomou et al. 2017 120): in bulk solutions, we have shown that due to their strong attractive forces, TEQ vesicles and CNCs associate and form large aggregates (> 1 µm). Within these aggregates, vesicles can be visualized using fluorescence microscopy. Furthermore, using wide-angle X-ray scattering as a function of temperature, it has been shown that the surfactants are in a hexagonal gel phase up to 60 ° C, this gel phase ensuring high stability of the vesicles in various conditions. As shown by ellipsometry, this mechanism ensures a uniform distribution of the surfactants on the cellulose surface, the layer being then a few nanometers thick. In this work, we also sought to test the hypothesis that the softness performance depends on the mass adsorbed at the cellulosic interfaces, and on the possibility that the guars improve this deposit. In the presence of guars, the adsorbed quantities measured by QCM-D are less (by 60%) than the sum of the respective amounts of each component, but the structure of the soft interface is more homogenous and rigid. This outcome suggests synergistic effects between surfactants and these particular polysaccharides, and that the guars could also contribute to the softness mechanism. These results finally show that surface techniques coupled with robust CNC coated substrates are promising for studying interactions of the many components of current conditioner formulations with cotton.

# Acknowledgements

G. M. L. Messina gratefully acknowledges the financial support by CSGI. ANR (Agence Nationale de la Recherche) and CGI (Commissariat à l'Investissement d'Avenir) are gratefully acknowledged for their financial support of this work through Labex SEAM (Science and Engineering for Advanced Materials and devices) ANR 11 LABX 086, ANR 11 IDEX 05 02. We acknowledge the ImagoSeine facility (Jacques Monod Institute, Paris, France), and the France BioImaging infrastructure supported by the French National Research Agency (ANR-10-INSB-04, « Investments for the future »). This research was supported in part by the Agence Nationale de la Recherche under the contract ANR-13-BS08-0015 (PANORAMA), ANR-12-CHEX-0011 (PULMONANO), ANR-15-CE18-0024-01 (ICONS), ANR-17-CE09-0017 (AlveolusMimics) and by Solvay.

# Conflict of Interest
The authors declare that they have no conflict of interest.

# Availability of data and material
Not applicable

# Code availability
Not applicable

# Graphic abstract

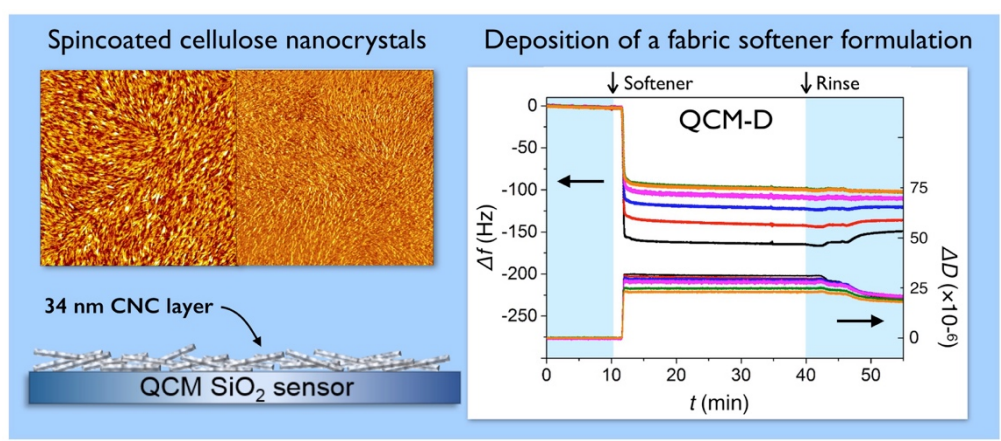